\documentclass[twocolumn,aps,prb,groupedaddress,showkeys,showpacs]{revtex4}
\usepackage{bm}
\usepackage{graphicx}
\catcode`@=11
\def\matrix#1{\null\,\vcenter{\normalbaselines\m@th
    \ialign{\hfil$##$\hfil&&\quad\hfil$##$\hfil\crcr
      \mathstrut\crcr\noalign{\kern-\baselineskip}
      #1\crcr\mathstrut\crcr\noalign{\kern-\baselineskip}}}\,}

\def\eqalign#1{\null\,\vcenter{\openup\jot\m@th
\ialign{\strut\hfil$\displaystyle{##}$&$\displaystyle{{}##}$\hfil
\crcr#1\crcr}}\,}
\def\Journal #1,#2,#3,#4#5#6#7{#1 {\bf #2}, #3 (#4#5#6#7)}
\def\lsim{\lower -0.3ex \hbox{$<$} \kern -0.75em \lower 0.7ex \hbox{$\sim$}}
\def\gsim{\lower -0.3ex \hbox{$>$} \kern -0.75em \lower 0.7ex \hbox{$\sim$}}
\catcode`@=12
\begin{document}
%
\pacs{73.63.Fg, 73.23.Ad, 72.80.Rj}
\keywords{carbon nanotube, double-wall carbon nanotube, inter-tube conductance, incommensurability, impurity}
%
\title{Electronic inter-tube transfer in double-wall carbon nanotubes with impurities}
\author{Seiji Uryu and Tsuneya Ando}
\affiliation{
Department of Physics, Tokyo Institute of Technology, 2-12-1 Ookayama, Meguro-ku, Tokyo 152-8551, Japan}
\date{\today}
%
\begin{abstract}
%
Inter-tube conductance of double-wall carbon nanotubes with impurities is numerically studied.
Its length dependence for various impurities is scaled by a mean-free path.
The inter-tube conductance exhibits drastic linear increase with the tube length and takes a maximum around at the localization length.
The maximum conductance is much smaller than the conductance quantum $e^2/\pi\hbar$.
\par
%
\end{abstract}
\maketitle
%
\section{Introduction} \label{Sec:Introduction}
%
Multi-wall carbon nanotubes are a coaxial multi-tube system which is realized by self-assembly.
It was theoretically revealed that electronic inter-tube transfer of the multi-wall nanotubes is negligibly small because of cancellation of inter-tube coupling among different sites due to quasi-periodicity of the system.
This implies that disorder disturbing this cancellation can play a crucial role in the inter-tube transfer.
In this paper inter-tube transfer of multi-wall carbon nanotubes with impurities is studied by numerical calculations for the simplest multi-wall nanotubes, i.e., double-wall nanotubes.
\par
%
The structure of double-wall tubes has two features.
One is that lattices of outer and inner tubes are incommensurate, leading to quasi-periodicity that the ratio between periods of the outer and inner tubes in a tube axis direction is irrational.\cite{Kociak_et_al_2002a,Zuo_et_al_2003a}
The other is that distance between the tubes is about $3.6 \pm 0.3$ {\AA}.\cite{Bandow_et_al_2001a}
Multi-wall tubes are considered to have the similar features.
\par
%
Using multi-wall nanotubes, inter-tube transfer has been experimentally studied.
In telescoping multi-wall tubes, inter-tube resistance was measured as a function of extraction distance of inner core tubes.\cite{Cumings_and_Zettle_2004a}
Results show monotonic increase with increase of the extraction distance.
It was also reported that in multi-terminal geometries nonlocal voltage drop occurs in the outer tube when current flows in the inner tube.\cite{Bourlon_et_al_2004a}
This was attributed to electronic inter-tube conduction and inter-tube conductivity was estimated.
\par
%
Inter-tube conduction has also been theoretically studied.\cite{Charlier_et_al_2007a}
It was revealed that the incommensurability of lattice leads to negligibly small inter-tube conductance as compared to that within each tube.\cite{Yoon_et_al_2002a,Triozon_et_al_2004a,Uryu_and_Ando_2005c}
On the other hand, there have been few theoretical studies on inter-tube transfer in the presence of impurities.
In some special cases such as tubes with a single impurity\cite{Tunney_and_Cooper_2006a} and short tubes with impurities,\cite{Uryu_and_Ando_2006a} enhancement of inter-tube conductance was shown.
In order to quantitatively understand disorder effects, we shall perform a systematic study of inter-tube conductance of double-wall tubes with impurities by numerical calculations.
\par
%
The paper is organized as follows:
In Sec.\ II the model and method are described.
Numerical results are presented in Sec.\ III and discussed in Sec.\ IV.
Summary and conclusion are given in Sec.\ V.
\par
%
\section{Model and Method} \label{Sec:Model_and_Method}
%
A tight-binding model with $\pi$ orbital is used.
A double-wall tube consists of an outer tube (tube 1) and an inner tube (tube 2).
We consider an intra-tube resonance integral $-\gamma_0$ between the nearest-neighbor sites and an inter-tube resonance integral $-t({\bf R}_1,{\bf R}_2)$ between sites ${\bf R}_1$ of tube 1 and ${\bf R}_2$ of tube 2.
The latter is chosen as\cite{Nakanishi_and_Ando_2001a,Uryu_2004a,Slater_and_Koster_1968a}
%
\begin{equation}
\eqalign{
&-t({\bf R}_1,{\bf R}_2) = V_{pp}^\sigma\exp\Big(-{d-c/2\over\delta}\Big)\Big({{\bf p}_1\!\cdot\!{\bf d}\over d}\Big)\Big({{\bf p}_2\!\cdot\!{\bf d}\over d}\Big)\cr
& \enspace -V_{pp}^\pi\exp\Big(-{d-a_0\over\delta}\Big)
\Big[({\bf p}_1\cdot{\bf e})({\bf p}_2\cdot{\bf e})+({\bf p}_1\cdot{\bf f})({\bf p}_2\cdot{\bf f})\Big], \cr
}
\label{Eq:Inter-tube_transfer}
\end{equation}
%
where $a_0$ is the distance between neighboring carbons in two-dimensional graphite given by $a_0/a\!=\!1/\sqrt{3}$ with lattice constant $a=2.46$ {\AA}, $c$ the lattice constant along the $c$ axis in bulk graphite given by $c/a\!=\!2.72$, and $\delta$ the decay rate of $\pi$ orbital.
Further, two parameters $V_{pp}^\sigma$ and $V_{pp}^\pi$ indicate resonance integrals between $\pi$ orbitals decomposed into the directions parallel and perpendicular, respectively, to the vector ${\bf d}={\bf R}_1-{\bf R}_2$.
Vectors ${\bf p}_1$ and ${\bf p}_2$ are unit vectors directed along $\pi$ orbitals at ${\bf R}_1$ and ${\bf R}_2$, respectively, $d=|{\bf d}|$, and ${\bf e}$ and ${\bf f}$ unit vectors perpendicular to ${\bf d}$ and to each other.
\par
%
In the following numerical calculations we use parameters $\delta/a\!=\!0.185$,\cite{Lambin_et_al_2000a} $V_{pp}^\pi=\gamma_0$, and $V_{pp}^\sigma/\gamma_0\!=\!0.164$.
The value of $V_{pp}^\sigma$ is chosen by fitting the energy dispersion of graphite in the $c$-axis direction calculated with the use of Eq.\ (\ref{Eq:Inter-tube_transfer}) to that in the effective model.\cite{McClure_1957a,Slonczewski_and_Weiss_1958a}
The calculated inter-tube conductances do not qualitatively depend on these values as long as the variation is small.
\par
%
The strength of inter-tube interaction is characterized by its maximum value $t_{\rm max}$ when ${\bf R}_2$ is located just below ${\bf R}_1$, given by
%
\begin{equation}
t_{\rm max} = V_{pp}^\sigma\exp\Big(-{\Delta r-c/2\over\delta}\Big) ,
\label{Eq:Maximum_inter-tube_transfer}
\end{equation}
%
where $\Delta r$ is the inter-tube distance.
For $\delta=0.185\times a\approx0.46$ {\AA}, the variation $\Delta r\sim3.6\pm0.3$ {\AA} gives rise to sizable dependence of the inter-tube resonance integral on the structure.
In fact, $t_{\rm max}/\gamma_0$ takes a value from 0.18 to 0.05 when $\Delta r$ changes from 3.3 to 3.9 {\AA}.
\par
%
Short-range impurities on tube $i$ are modeled by random on-site potential which takes a value of $u_i$ or $-u_i$ with the same probability.
Using the Boltzmann transport equation in the Born approximation, a mean-free path for channel $\nu$ of tube $i$ is given by\cite{Akera_and_Ando_1991a,Seri_and_Ando_1997a}
%
\begin{equation}
{1\over l_{i,\nu}}={\sqrt{3}\over4L_i}\Bigl({au_i\over\hbar}\Bigr)^2\sum_{\mu}{2\over|v_{i,\nu}v_{i,\mu}|},
\label{Eq:Mean-free_path_nu}
\end{equation}
%
where $L_i$ is the circumference of tube $i$, the sum runs over channel $\mu$, and $v_{i,\mu}$ is the group velocity for channel $\mu$ of tube $i$.
We define an averaged mean-free path over outer and inner tubes as
%
\begin{equation}
l_e={N_1l_1+N_2l_2\over N_1+N_2},
\label{Eq:Mean-free_path}
\end{equation}
%
where $N_i$ is the number of channels in tube $i$ and $l_i$ a mean-free path of tube $i$ averaged over channels.
Because $l_{i,\nu}$ additively contributes to the conductivity,\cite{Akera_and_Ando_1991a,Seri_and_Ando_1997a} we shall define $l_i$ as
%
\begin{equation}
l_{i}={1\over N_{i}}\sum_{\nu}l_{i,\nu} .
\label{Eq:Mean-free_path_i}
\end{equation}
%
From Eq.\ (\ref{Eq:Mean-free_path_nu}), the mean-free path is longer in the outer tube than in the inner tube when magnitude of impurity potential is same in the two tubes.
Although there are other possible definitions of averaging, the following results are almost independent of details in the definition.
\par
%
\begin{figure}
\begin{center}
\includegraphics[width=7.0cm]{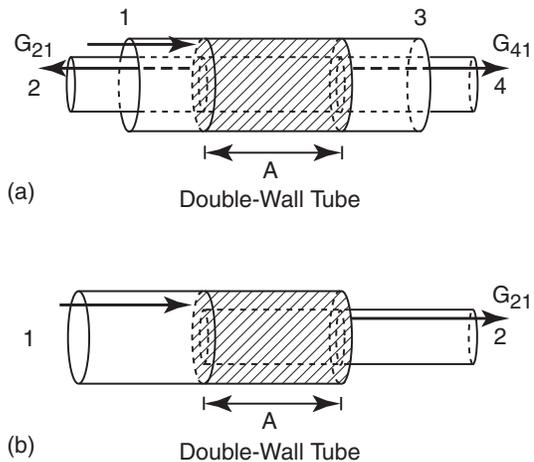}
\caption{Schematic illustrations of (a) a four-terminal tube and (b) a two-terminal tube.
Inter-tube transfer is considered only in the hatched double-wall region with length $A$.
Each terminal is numbered.
Arrows in (a) indicate current flow for the inter-tube conductance $G_{41}$ and $G_{21}$ and those in (b) indicate that for $G_{21}$.}
\label{Fig:4terminal_system}
\end{center}
\end{figure}
%
In this paper we consider double-wall tubes with metallic outer and inner tubes.
The conductance between the inner and outer tubes due to inter-tube transfer is calculated for two different types of the geometry.
One is a four-terminal system illustrated in Fig.\ \ref{Fig:4terminal_system}(a) and the other is a two-terminal system in Fig.\ \ref{Fig:4terminal_system}(b).
In the former system, inter-tube transfer is present only in the hatched double-wall region with length $A$, while tubes are independent outside the region and connected to reservoirs.
The four-terminal system is very advantageous to theoretical analysis of effects of inter-tube transfer itself because of absence of tube ends which can play significant roles in inter-tube transfer.\cite{Uryu_and_Ando_2005c,Uryu_and_Ando_2006a}
The two-terminal system is similar to telescoping tubes used in experiments.\cite{Cumings_and_Zettle_2004a}
\par
%
Using Landauer's formula,\cite{Landauer_1957a} the conductance $G_{\beta\alpha}$ for injection from terminal $\alpha$ into $\beta$ is defined by
%
\begin{equation}
G_{\beta\alpha}={e^2\over\pi\hbar}\sum_{\mu,\nu}|S_{\beta,\nu;\alpha,\mu}|^2,
\label{Eq:Landauer_formula}
\end{equation}
%
where $S_{\beta,\nu;\alpha,\mu}$ is the scattering matrix describing transmission from channel $\mu$ of terminal $\alpha$ to channel $\nu$ of terminal $\beta$ and is numerically calculated in a recursive Green function method.\cite{Ando_1991a}
For electron injection from the left side of outer tube as shown by arrows in Fig.\ \ref{Fig:4terminal_system}, the inter-tube conductances become $G_{41}$ and $G_{21}$ for the four-terminal tube and $G_{21}$ for the two-terminal tube.
For the four-terminal tubes $G_{41}$ and $G_{21}$ are called the inter-tube transmission and reflection conductance, respectively, in the following.
Similarly, the inter-tube transmission and reflection conductances for injection from terminal 2 are $G_{32}$ and $G_{12}$, respectively.
\par
%
Since the inter-tube resonance integral is small, the inter-tube conductance can be given in the lowest-order approximation.\cite{Uryu_and_Ando_2005c}
Define an effective inter-tube coupling at ${\bf R}_1$ on tube 1 as
%
\begin{equation}
t^{\mu\nu}({\bf R}_1) = \sum_{{\bf R}_2} \Psi_1^{\mu}({\bf R}_1)^*t({\bf R}_1,{\bf R}_2) \Psi_2^{\nu}({\bf R}_2) ,
\label{Eq:Effective_inter-tube_coupling}
\end{equation}
%
where $\Psi_1^{\mu}({\bf R}_1)$ and $\Psi_2^{\nu}({\bf R}_2)$ are the wave function in the absence of inter-tube transfer for channel $\mu$ of tube 1 and for channel $\nu$ of tube 2, respectively, and thus are plane waves.
The effective inter-tube coupling at ${\bf R}_2$ on tube 2, $t^{\mu\nu}({\bf R}_2)$, is given by Eq.\ (\ref{Eq:Effective_inter-tube_coupling}) with exchange of the subscripts 1 and 2.
\par
%
Because the phase of the wave functions around the K and K' point changes by $\pm2\pi/3$ for the position change by the primitive lattice vector and lattices of outer and inner tubes are incommensurate, this effective inter-tube coupling quasi-periodically depends on the position ${\bf R}_1$.
Therefore, they cancel each other when summed up over all the sites of the outer tube.
Small inter-tube conductance can appear by incomplete cancellation due to sudden truncation of the inter-tube resonance integrals at boundaries of finite tubes.
Therefore, when the inter-tube resonance integrals are switched on and off near the boundaries smoothly, the cancellation becomes complete and the inter-tube conductance becomes negligibly small.
\par
%
\begin{figure*}[t]
\begin{center}
\includegraphics[width=8.0cm]{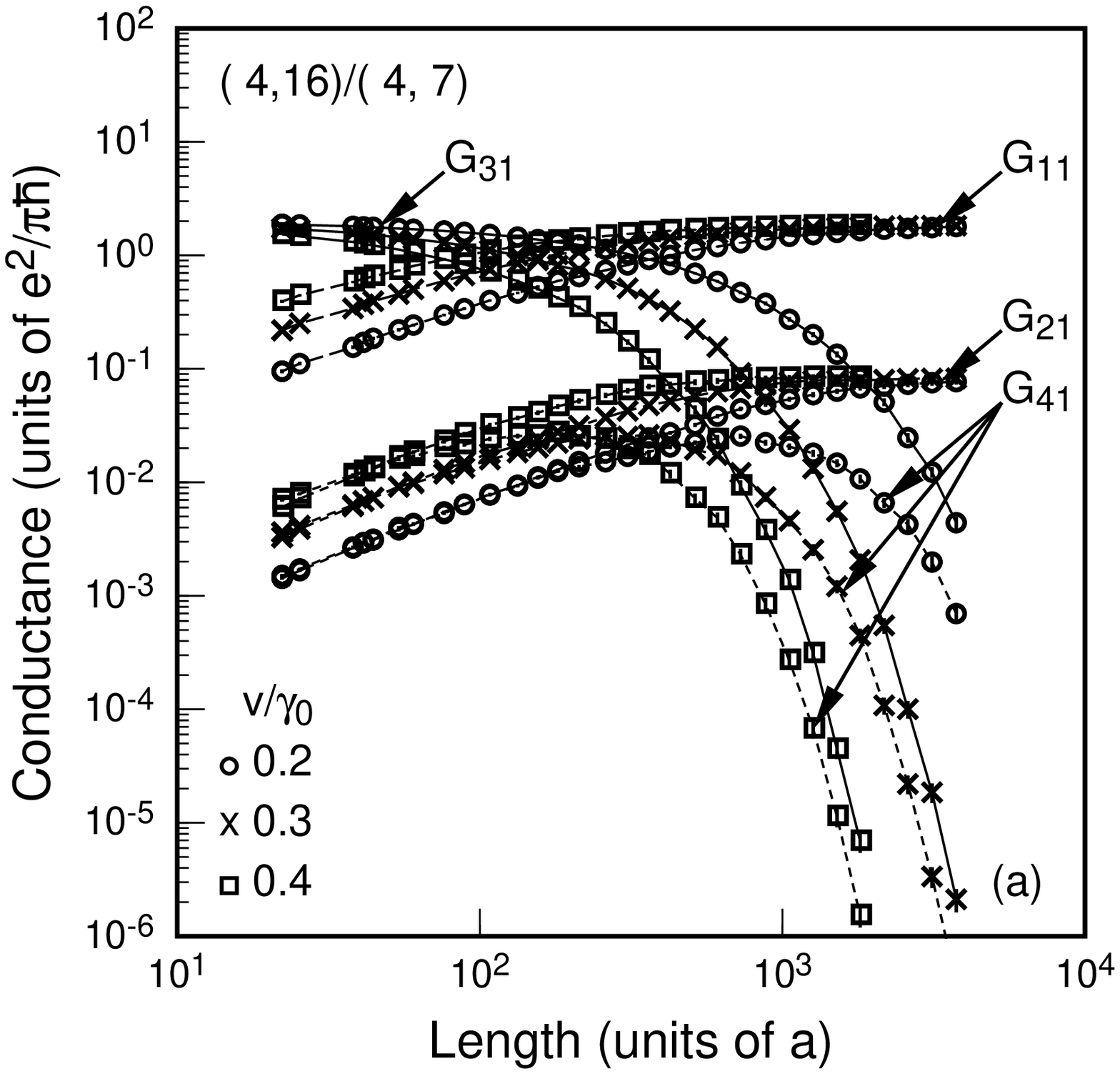}
\includegraphics[width=8.0cm]{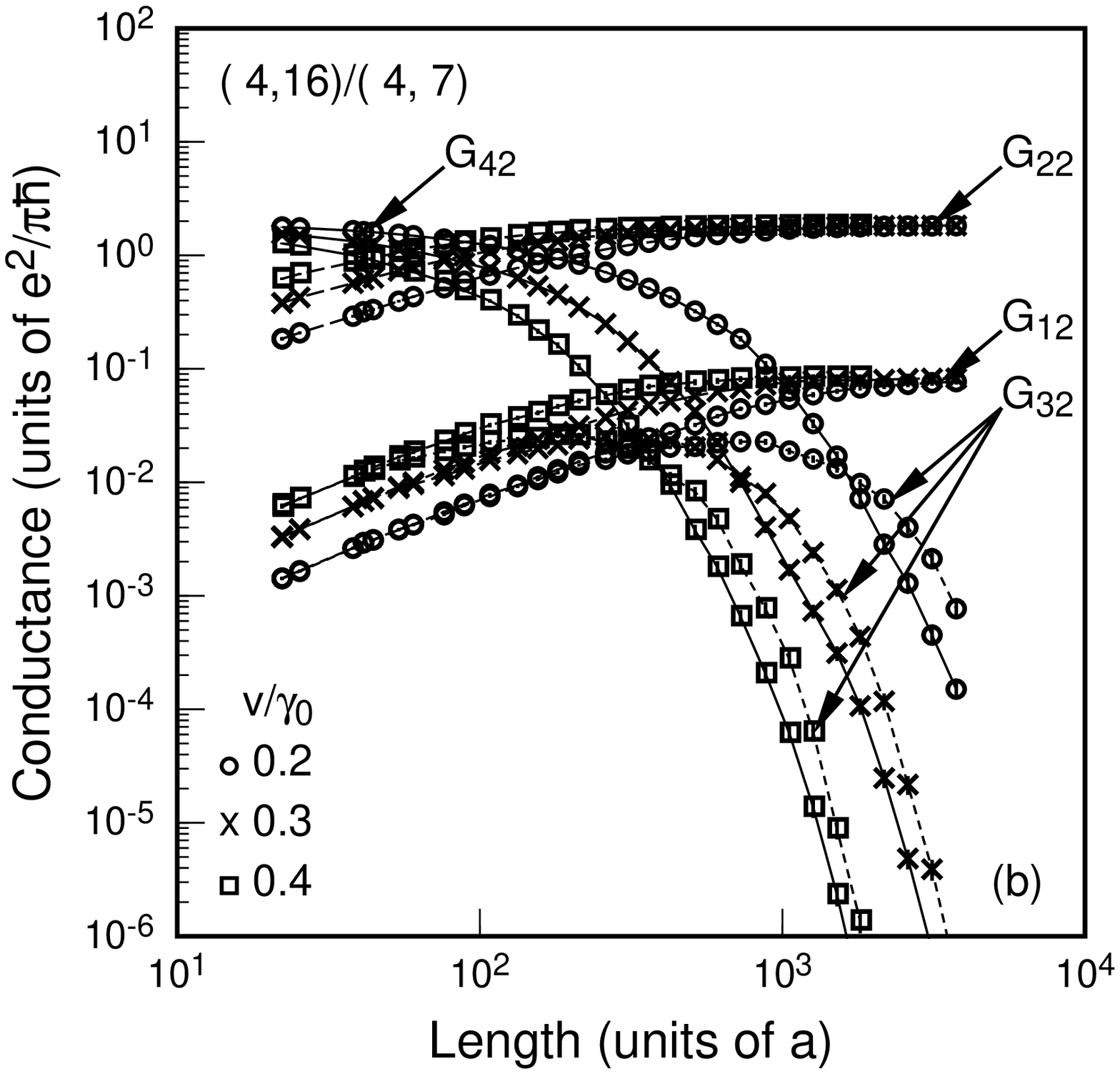}
\caption{Length dependence of conductances of a (4,16)/(4,7) tube.
The length is in units of the lattice constant.
Circles are for $u/\gamma_0=0.2$, crosses 0.3, and squares 0.4.}
\label{Fig:Conductance_of_A}
\end{center}
\begin{center}
\includegraphics[width=8.0cm]{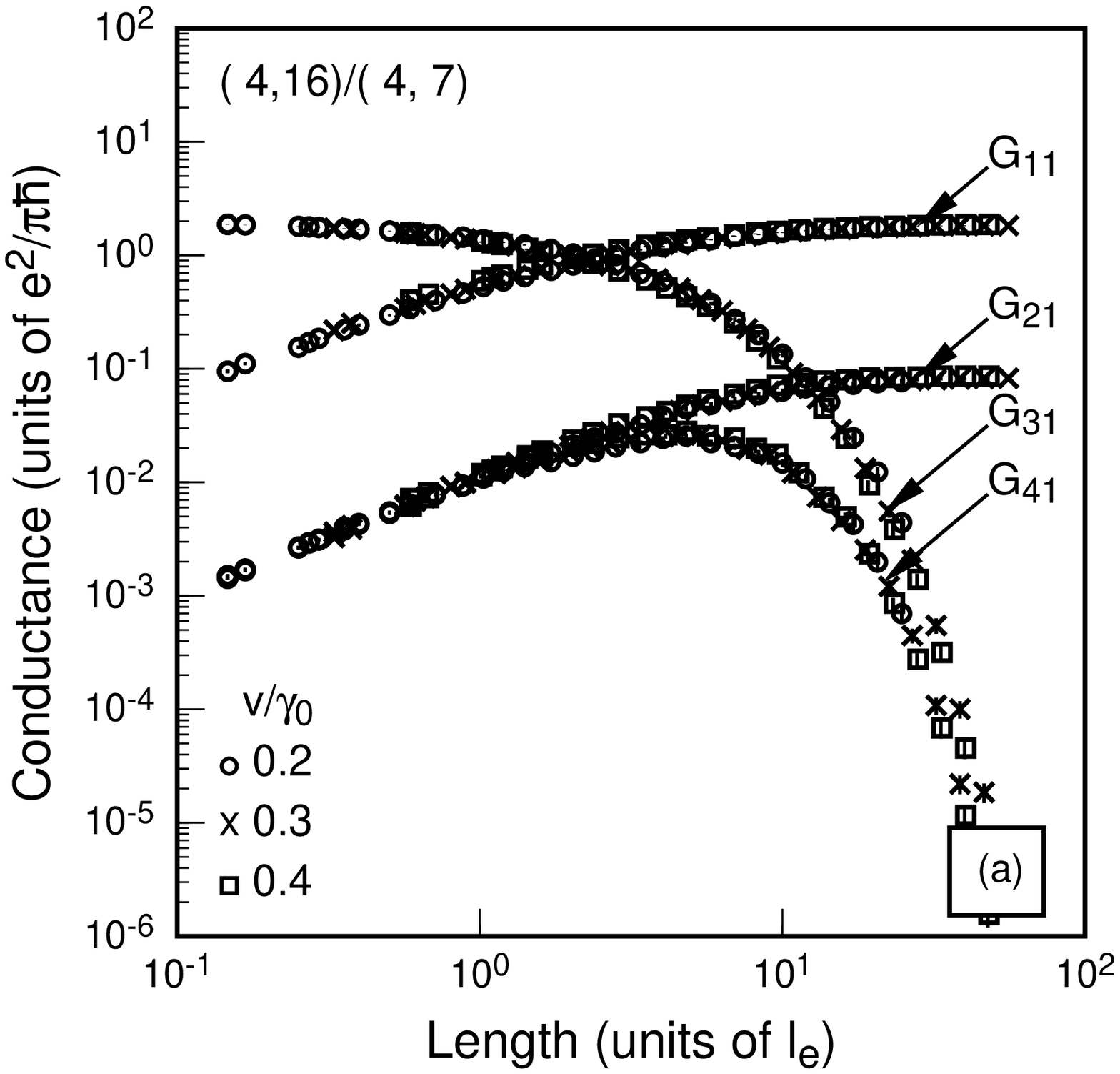}
\includegraphics[width=8.0cm]{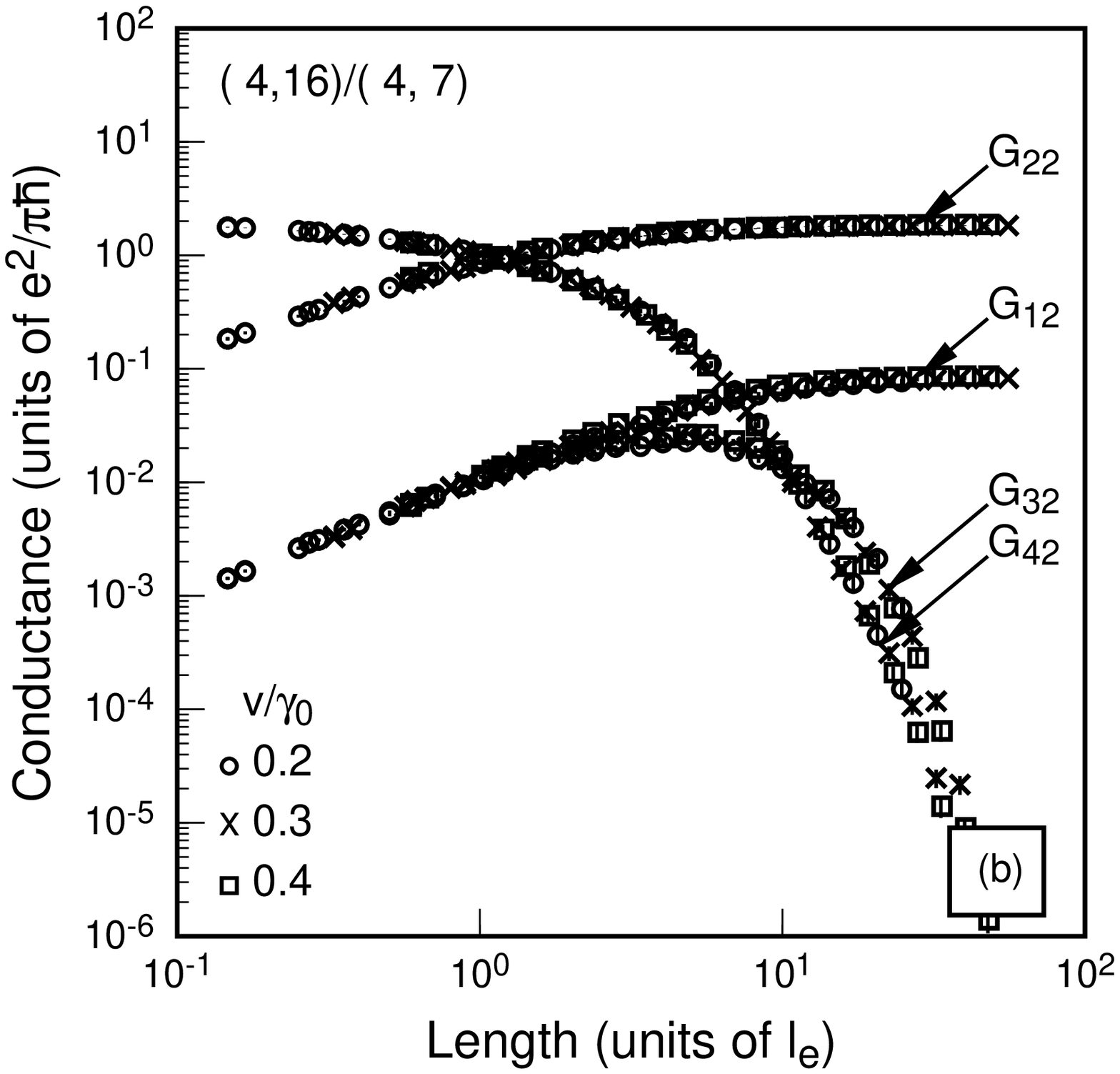}
\caption{Length dependence of conductances of a (4,16)/(4,7) tube.
The length is measured in units of the mean-free path $l_e$.
Circles are for $u/\gamma_0=0.2$, crosses 0.3, and squares 0.4.}
\label{Fig:Conductance_scaling}
\end{center}
\end{figure*}
%
\begin{figure}
\begin{center}
\includegraphics[width=8.0cm]{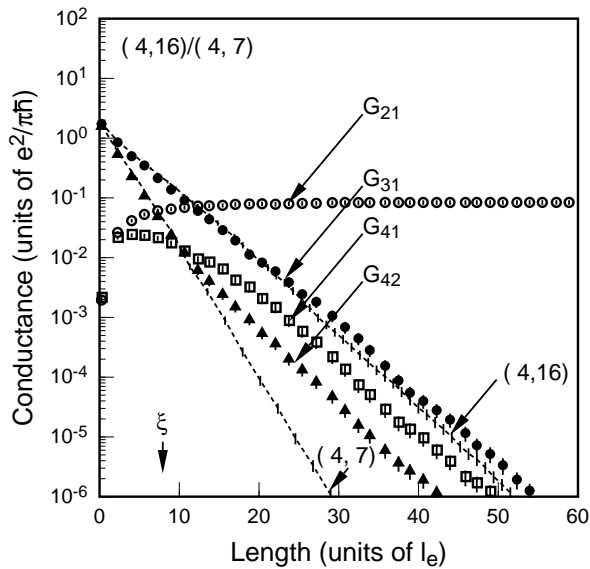}
\caption{Length dependence of various conductances of a (4,16)/(4,7) tube.
Closed circles and triangles are the intra-tube conductances $G_{31}$ and $G_{42}$, respectively, and open squares and circles the inter-tube conductances $G_{41}$ and $G_{21}$, respectively.
A dashed and dotted line denotes the conductance of single-wall (4,16) and (4,7) tube, respectively.
A vertical arrow indicates a localization length $\xi$.}
\label{Fig:Conductance_for_long_tube}
\end{center}
\end{figure}
%
Because of the time reversal symmetry, there is a relation for both the four- and two-terminal tubes
%
\begin{equation}
G_{\beta\alpha}=G_{\alpha\beta} .
\label{Eq:Time-reversal}
\end{equation}
%
Under the condition that the inter-tube conductance is dominated by impurities, the system has the mirror reflection with respect to the plane perpendicular to the axis.
Then, there is a relation between the inter-tube transmission conductances of the four-terminal tube
%
\begin{equation}
G_{41}=G_{23} .
\label{Eq:Reflection_symmetry_1}
\end{equation}
%
With the use of Eq.\ (\ref{Eq:Time-reversal}), this leads to
%
\begin{equation}
G_{41}=G_{32}.
\label{Eq:G41=G32}
\end{equation}
%
\par
%
As mentioned before, variation of the inter-tube distance affects the inter-tube conductance because of non-negligible variation of the inter-tube coupling strength.
Since the inter-tube conductance is well described in the lowest-order approximation,\cite{Uryu_and_Ando_2005c} such an effect is expected to be factored out by dividing the inter-tube conductance by absolute square of the inter-tube coupling strength.
Representing the strength by the maximum value of inter-tube resonance integral in Eq.\ (\ref{Eq:Maximum_inter-tube_transfer}), we define a normalized inter-tube conductance $\tilde G_{\beta\alpha}$ as
%
\begin{equation}
G_{\beta\alpha}=\tilde G_{\beta\alpha}\Bigl({t_{\rm max}\over\gamma_0}\Bigl)^2 ,
\label{Eq:Normalized_conductance}
\end{equation}
%
with $(\beta,\alpha)=(4,1)$, (2,1), etc.
\par
%
The strength of impurity potential will be chosen as the same between outer and inner tubes, i.e., $u_1=u_2=u$, unless otherwise specified.
The potential strength $|u|$ should be chosen to be weaker than $\gamma_0$.
In fact, effects of scattering exhibit qualitative change from the weak-scatterer regime for $|u|<\gamma_0$ to the strong-scattering regime like lattice vacancies for $|u|>\gamma_0$.\cite{Nakanishi_and_Ando_1999a,Ando_et_al_1999b}
In the latter case effects of scattering change dramatically depending on the impurity configuration.\cite{Ando_et_al_1999b,Igami_et_al_TB_All}
\par
%
In the following, results for a (4,16)/(4,7) tube, with outer (4,16) and inner (4,7), are shown as typical examples unless specified otherwise.
The energy is set to be close to zero, $E(\sqrt{3}\pi a\gamma_0/L_1)^{-1}=-0.05$, where $\sqrt{3}\pi a\gamma_0/L_1$ is the bottom of the first excited subband for sufficiently thick outer nanotubes.
This gives $N_1=N_2=2$.
In order to make the inter-tube conductance without impurities negligible we use the inter-tube resonance integrals which is smoothly switched on and off near the tube edges.\cite{Uryu_and_Ando_2005c,Uryu_and_Ando_2006a}
Results do not qualitatively depend on details of the above.
We show the geometric average of the conductances over impurity configurations which will be denoted as $G_{\beta\alpha}$ for simplicity.
\par
%
\section{Numerical Results} \label{Sec:Numerical_Results}
\subsection{Four-terminal system} \label{Ssc:Four-terminal_system}
%
Figure \ref{Fig:Conductance_of_A} shows length dependence of conductances for electron injection from (a) terminal 1 of the outer tube and (b) terminal 2 of the inner tube for a four-terminal (4,16)/(4,7) tube.
The length is plotted in units of the lattice constant.
The circles denote results for $u/\gamma_0=0.2$, the crosses for 0.3, and the squares 0.4.
The results are strongly dependent on this disorder parameter.
\par
%
When the length is scaled by the mean-free path $l_e$ given by Eq.\ (\ref{Eq:Mean-free_path}), each conductance in Figs.\ \ref{Fig:Conductance_of_A} (a) and (b) collapses onto a single curve as shown in Figs.\ \ref{Fig:Conductance_scaling} (a) and (b).
For the length shorter than the mean-free path, the inter-tube transmission and reflection conductances are the same, i.e., $G_{41}=G_{21}=G_{12}=G_{32}$, and linearly increase with the length.
The inter-tube conductance takes a maximum around the length several times as long as the mean-free path.
For tubes much longer than the mean-free path, transmission conductances $G_{41}$ and $G_{32}$ decrease and reflection conductances $G_{21}$ and $G_{12}$ are saturated because the electron wave function starts to decrease exponentially due to localization effects caused by the disorder
\par
%
\begin{figure*}[t]
\begin{center}
\includegraphics[width=8.0cm]{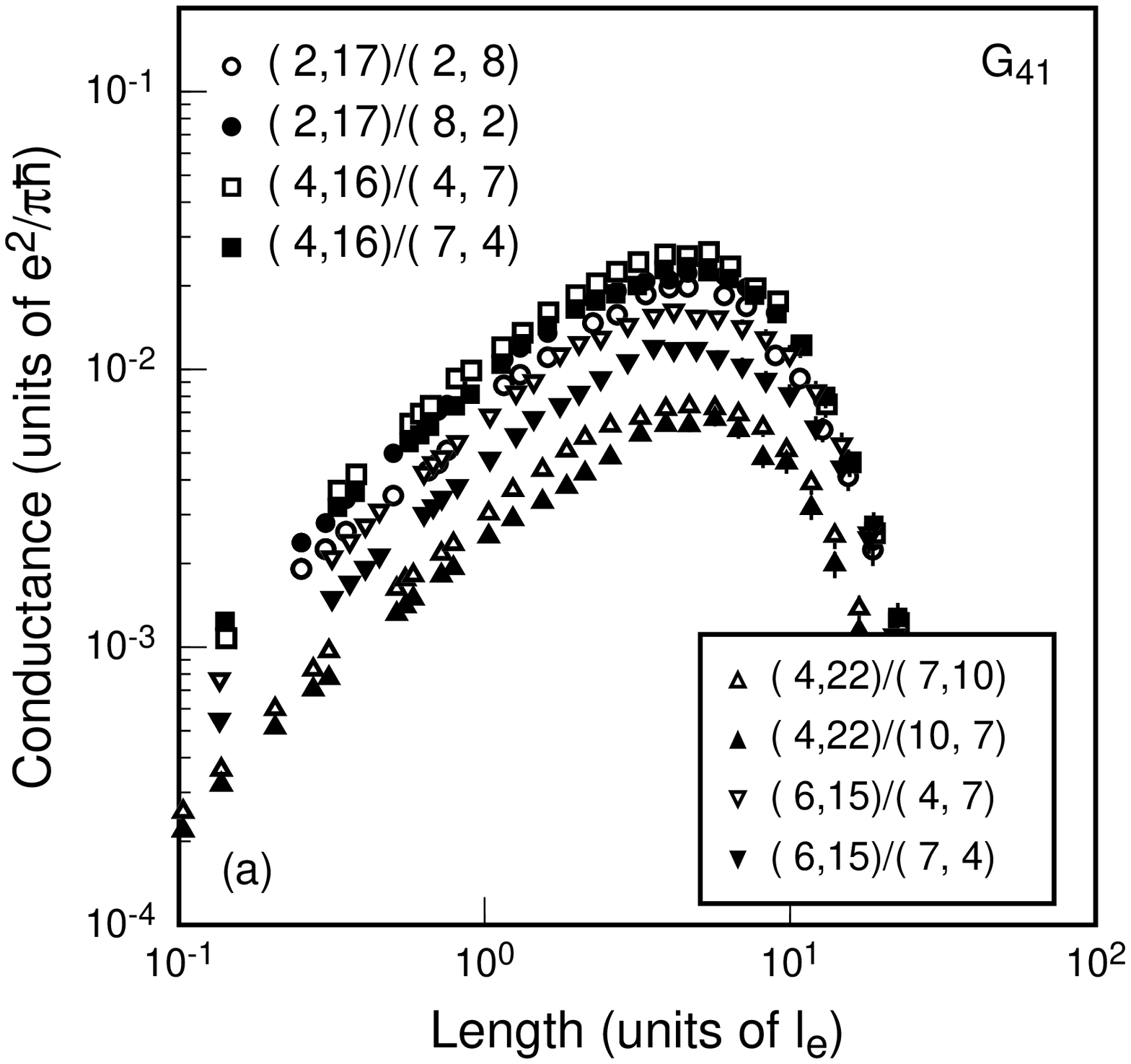}
\includegraphics[width=8.0cm]{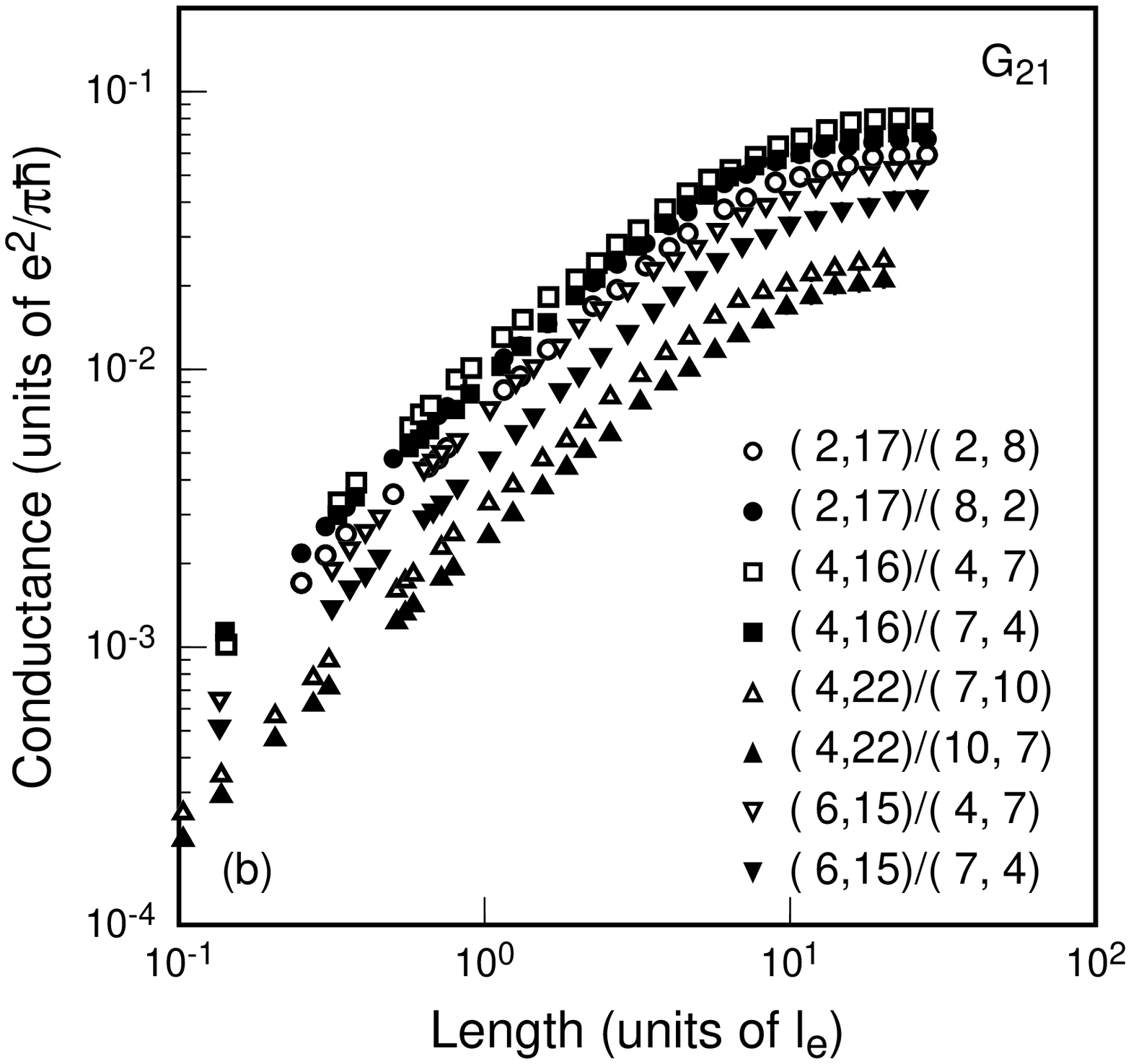}
\caption{Length dependence of the inter-tube (a) transmission and (b) reflection conductances for various tubes.}
\label{Fig:Various_structures}
\end{center}
\begin{center}
\includegraphics[width=8.0cm]{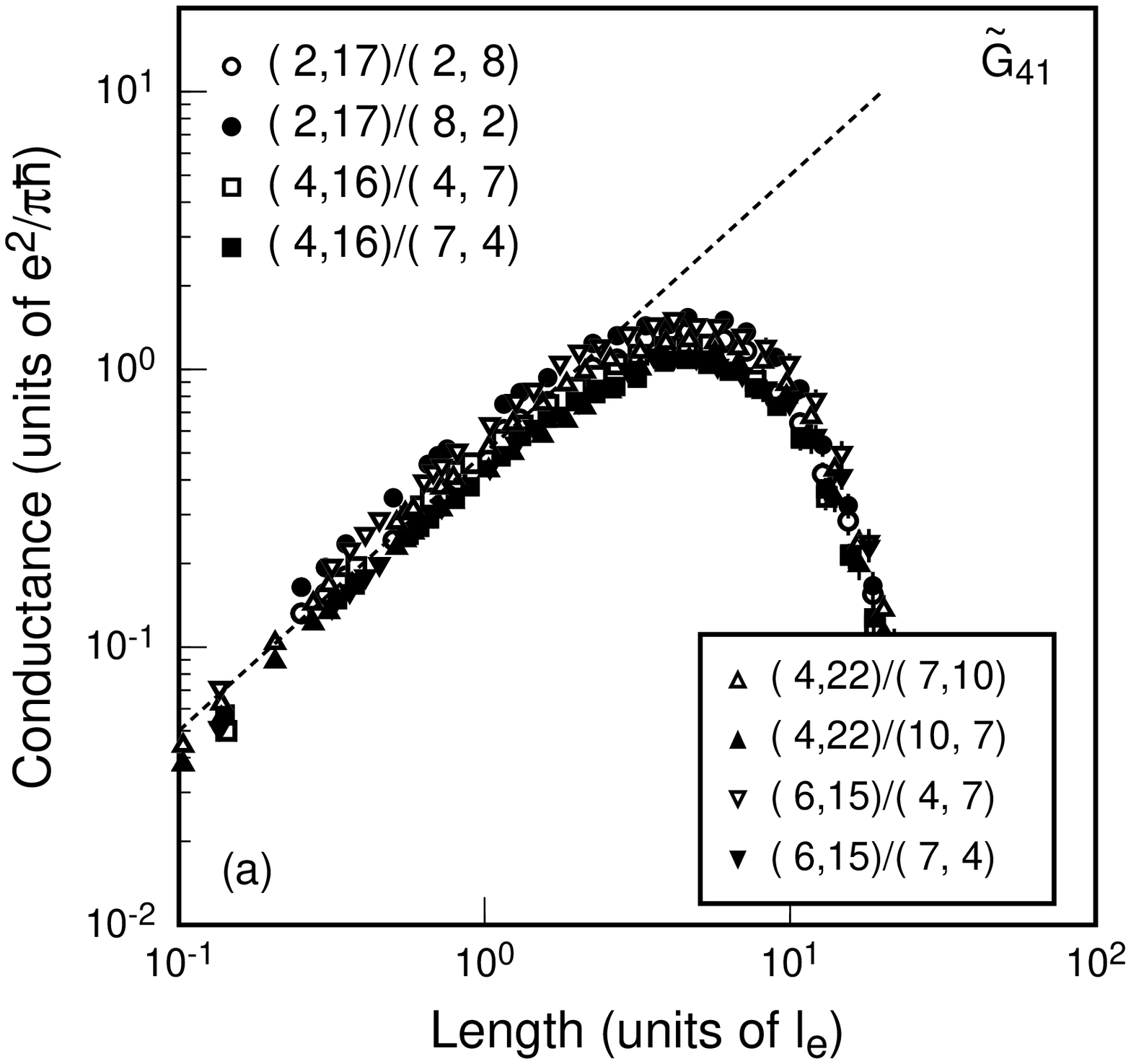}
\includegraphics[width=8.0cm]{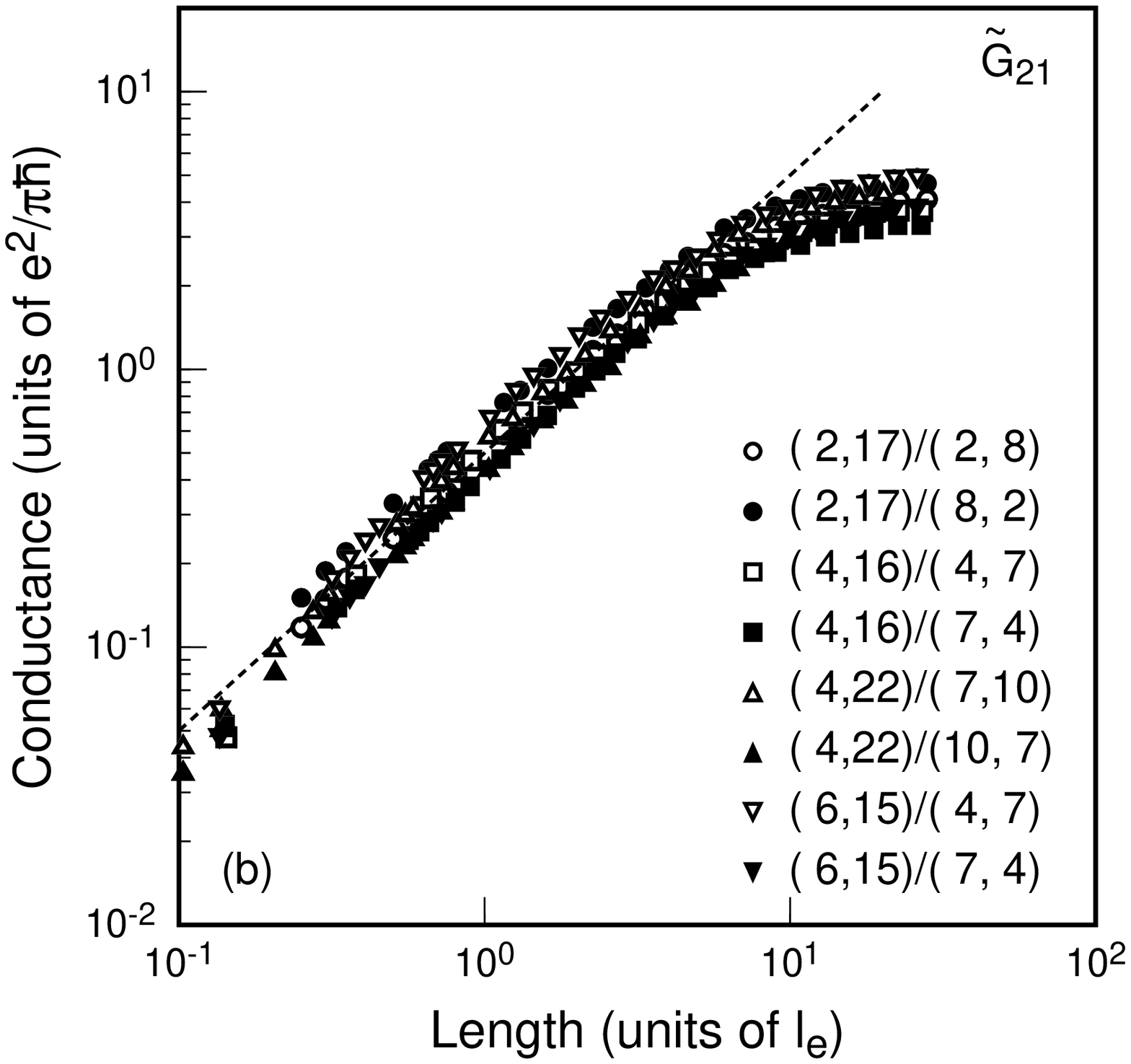}
\caption{Length dependence of the normalized inter-tube (a) transmission and (b) reflection conductances for various tubes.
Dotted lines indicate $0.5\times (e^2/\pi\hbar)(A/l_e)$.}
\label{Fig:Normalized_conductance}
\end{center}
\end{figure*}
%
\begin{figure}
\begin{center}
\includegraphics[width=8.0cm]{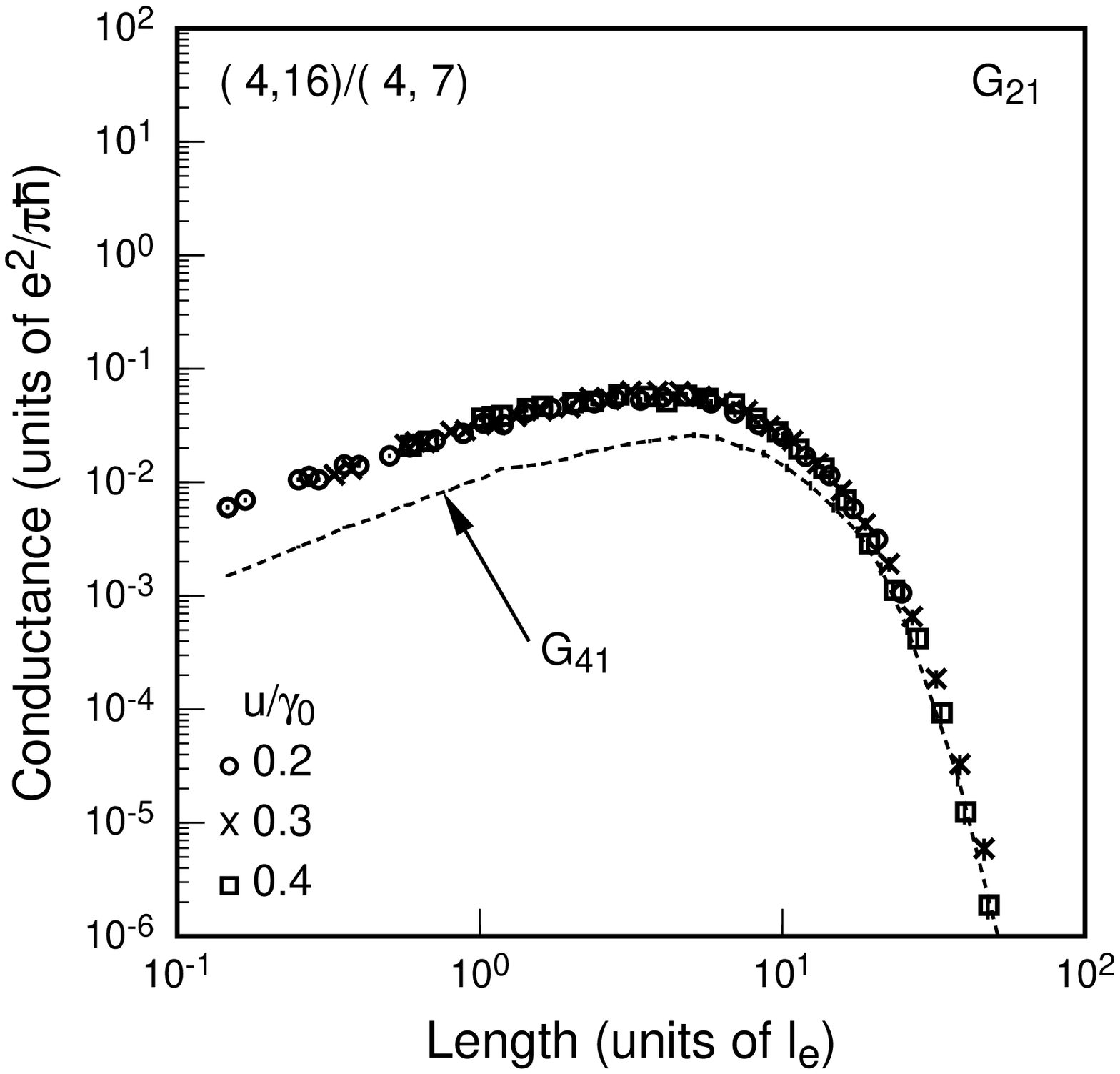}
\caption{Length dependence of the inter-tube conductance for a two-terminal (4,16)/(4,7) tube.
Circles are for $u/\gamma_0=0.2$, crosses 0.3, and squares 0.4.
A dotted line indicates the inter-tube transmission conductance $G_{41}$ of a four-terminal (4,16)/(4,7) tube.}
\label{Fig:Two-terminal_conductance}
\end{center}
\end{figure}
%
The linear length dependence arises because impurities destroy the quasi-periodic position dependence of the effective inter-tube couplings and disturb their cancellation.
The equality between the inter-tube transmission and reflection conductances $G_{41}=G_{21}=G_{12}=G_{32}$ for short tubes arises from the fact that electrons transferred between the tubes go right or left with the same probability.
\par
%
In Fig.\ \ref{Fig:Conductance_for_long_tube} the intra-tube transmission conductances for the outer and inner tubes $G_{31}$ and $G_{42}$, respectively, and the inter-tube conductances $G_{41}$ and $G_{21}$ are plotted as a function of length measured in units of $l_e$.
Dashed and dotted lines are conductances of single-wall (4,16) and (4,7) tubes, respectively, with the same impurities.
Since the mean-free path of the (4,16) tube is longer than that of the (4,7) tube, the localization length of the (4,16) tube is larger than that of the (4,7) tube, also.
\par
%
The intra-tube conductance $G_{31}$ exponentially decays and the localization length $\xi$ defined by $G_{31}\propto\exp(-2A/\xi)$ is close to but slightly (twelve percent)  longer than that of the single-wall (4,16) tube, which is $\xi/l_e\approx8$ as shown by a vertical arrow.
The intra-tube conductance $G_{42}$ decays with the localization length of the single-wall (4,7) tube for $A/l_e\lsim10$ and changes to decrease with that of $G_{31}$ for $A/l_e\gsim10$.
The inter-tube transmission conductance $G_{41}$ exponentially decays with the same localization length $\xi$ at sufficiently large length.
This indicates that localization length of adequately long double-wall tubes is approximately determined by that of the constituent tube with a longer mean-free path.
It should be noted that maximum values of the inter-tube conductances reaches approximately at $A\sim\xi$, although saturation of their linear increase starts at a length shorter than $\xi$.
\par
%
In metallic single-wall nanotubes, the backscattering is suppressed for scatterers with potential range larger than or comparable to the lattice constant.\cite{Ando_and_Nakanishi_1998a,Ando_et_al_1998b}
This is a result of the special symmetry present in the effective-mass equation equivalent to Weyl's equation or the Dirac equation with vanishing mass.\cite{Ando_and_Suzuura_2002a,Ando_2005a}
This symmetry is destroyed by various perturbations and short-range scatterers are one of them.
In the presence of such perturbations electrons are localized due to backscattering.\cite{Ando_2004b,Ando_and_Akimoto_2004a,Akimoto_and_Ando_2004a,Ando_2006c}
The present localization length is also written as $\xi/l_1\approx6$ when measured in units of the averaged mean-free path of the outer tube.
This value is approximately the same as that previously calculated for dominant short-range scatterers.\cite{Ando_and_Akimoto_2004a}
However, it is larger than $\xi/l_e\approx1.5$ reported in Ref.\ \onlinecite{Avriller_et_al_2006a}, presumably due to the difference in the definition of the localization length and mean-free path.
\par
%
Figures \ref{Fig:Various_structures}(a) and (b) show the inter-tube conductance of transmission $G_{41}$ and reflection $G_{21}$, respectively, for eight different tubes where the inter-tube distance $(L_1-L_2)/2\pi$ ranges from $\sim3.4$ to $\sim3.7$ {\AA}.
It can be seen that the conductance strongly depends on the tube structure.
For example, the largest conductance of the (4,16)/(4,7) tube (open squares) is about four times as large as the smallest of the (4,22)/(10,7) tube (closed triangles).
The normalized inter-tube conductances $\tilde G_{41}$ and $\tilde G_{21}$ given by Eq.\ (\ref{Eq:Normalized_conductance}) are shown in Figs.\ \ref{Fig:Normalized_conductance}(a) and (b), respectively, for the same tubes in Fig.\ \ref{Fig:Various_structures}.
Their structure dependence is clearly reduced comparing to that in Fig.\ \ref{Fig:Various_structures}.
\par
%
In the region where the conductance linearly depends on the length, we approximately have 
%
\begin{equation}
{\tilde G}_{41}\sim{\tilde G}_{21}\sim0.5\times{e^2\over\pi\hbar}{A\over l_e} ,
\label{Eq:A-linear_normalized_conductance}
\end{equation}
%
which is shown by dotted lines in Fig.\ \ref{Fig:Normalized_conductance}.
The maximum of the inter-tube conductance can be estimated by that at about the localization length $\xi=Cl_e$ with $C\approx8$ to be very roughly $0.5\times C(t_{\rm max}/\gamma_0)^2$ in units of $e^2/\pi\hbar$ from Eq.\ (\ref{Eq:Normalized_conductance}).
For example, it becomes $\sim0.13$ for $\Delta r=3.3$ {\AA}, $\sim0.03$ for 3.6 {\AA}, and $\sim0.009$ for 3.9 {\AA} in units of $e^2/\pi\hbar$.
The inter-tube conductance can be substantially enhanced but is small compared to the conductance quantum.
\par
%
It can be seen that there still remains small structure dependence in Fig.\ \ref{Fig:Normalized_conductance}.
For a fixed outer tube, the inter-tube distance for the inner $(n,m)$ tube is equal to that for the inner $(m,n)$ tube.
In Fig.\ \ref{Fig:Various_structures} four kinds of such pair tubes with the same inter-tube distance are indicated by open and closed symbols with the same shape.
The conductances for open and closed symbols with the same shape are not exactly equal to each other.
One reason for that is that the way of cancellation of the inter-tube couplings in the absence of impurities depends on the structure and therefore effects of impurities on disturbance of the cancellation also depend on the structure.
\par
%
\subsection{Two-terminal system} \label{Ssc:Two-terminal_system}
%
The length dependence of the inter-tube conductance $G_{21}$ for a two-terminal (4,16)/(4,7) tube is shown in Fig.\ \ref{Fig:Two-terminal_conductance}.
The conductances for three kinds of impurities are plotted as a function of the length in units of the mean-free path.
The qualitative features are same as the inter-tube transmission conductance of the four-terminal (4,16)/(4,7) tube.
For comparison, $G_{41}$ in Fig.\ \ref{Fig:Conductance_scaling}(a) is depicted by a dotted line.
\par
%
The conductance for a short length two-terminal tube is about four times as large as that of the corresponding four-terminal tube.
This factor four comes from the fact that almost all injected electrons are reflected at the tube end and the effective length becomes twice as long as the actual length.
On the other hand, the localization length obtained from $G_{21}$ is the same as that of the four-terminal tube, because the reflection at the edge does not play a role due to exponentially small amplitude there due to localization effects.
\par
%
\begin{figure*}[t]
\begin{center}
\includegraphics[width=8.0cm]{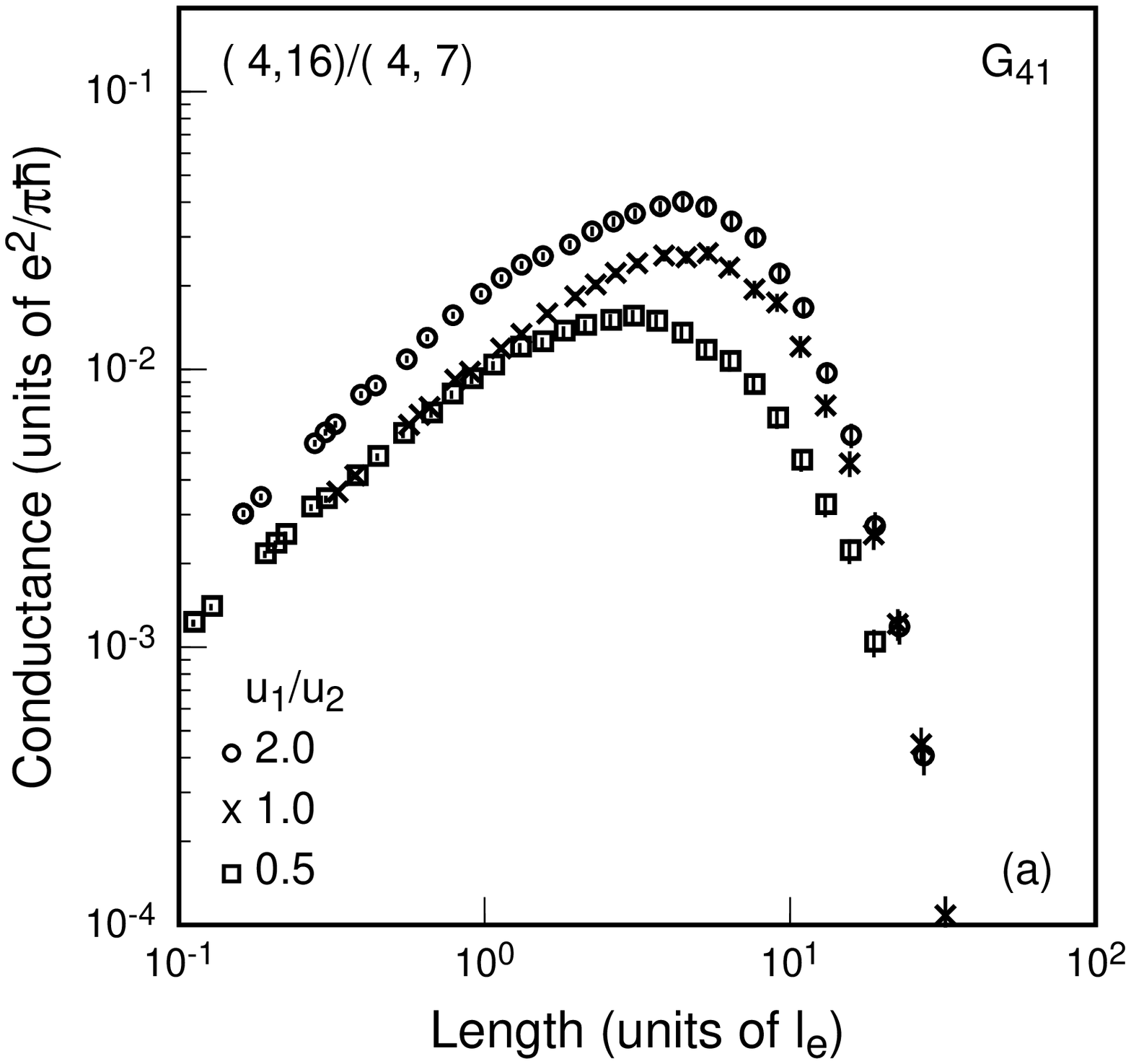}
\includegraphics[width=8.0cm]{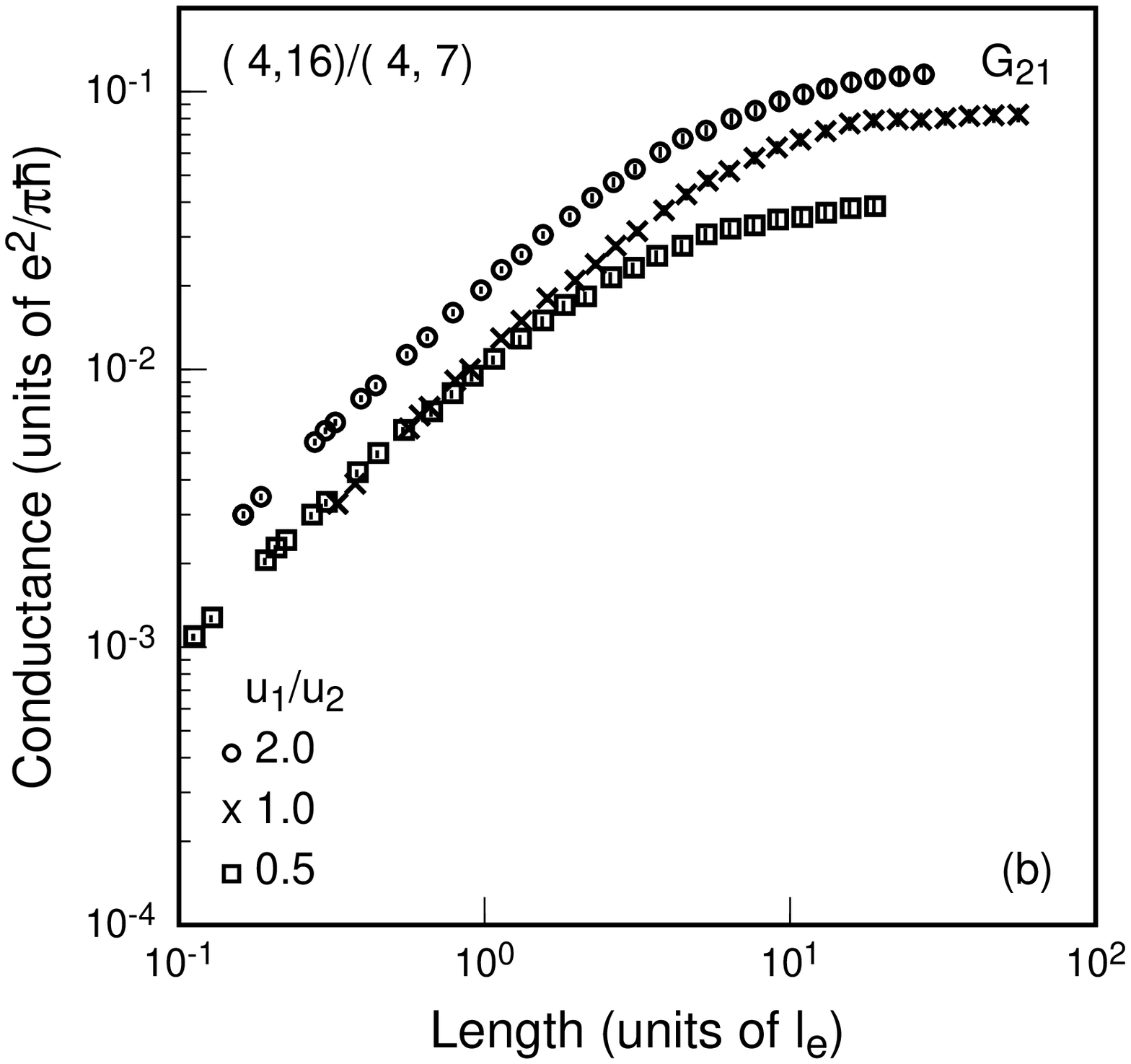}
\caption{Length dependence of the inter-tube (a) transmission and (b) reflection conductances for a four-terminal (4,16)/(4,7) tube.
Circles are for $u_1/u_2=2$, crosses 1, and squares 0.5.}
\label{Fig:Different_impurities}
\end{center}
\end{figure*}
%
\section{Discussions} \label{Sec:Discussions}
%
We shall first consider double-wall tubes consisting of outer and inner tubes with different disorder, i.e., $u_1\ne u_2$.
For tubes with a fixed ratio $u_1/u_2$, the same scaling of the conductances as the previous one holds.
Figures \ref{Fig:Different_impurities}(a) and (b) show length dependence of inter-tube transmission and reflection conductances, respectively, for $u_1/u_2=2$ (circles), 1 (crosses), and 0.5 (squares).
The mean-free paths of the outer and inner tubes are $l_1/l_2=0.5$ for $u_1/u_2=2$, 2 for $u_1/u_2=1$, and 8 for $u_1/u_2=0.5$.
The qualitative behavior of the results is the same in spite of the large variation in the mean free path.
Quantitatively, however, small differences exist in both the absolute value and the length dependence.
\par
%
This difference arises due to the asymmetry of disorder effects between the inner and outer tubes.
In fact the magnitude of the effective inter-tube couplings is larger for the outer tube, $t({\bf R}_1)$, than for the inner tube, $t({\bf R}_2)$, as an average.
This is because the number of sites of the outer tube connected to each site of the inner tube is larger due to the curvature, leading to more cancellation for $t({\bf R}_2)$.
In the absence of disorder, this difference of $t({\bf R}_1)$ and $t({\bf R}_2)$ is meaningless because the conductance is given by the sum of each, which are certainly the same.
The inter-tube conductance is induced by random modulation of the wave function due to impurities.
The resulting conductance is larger for the outer tube with larger $t({\bf R}_1)$ than for the inner tube with smaller $t({\bf R}_2)$.
\par
%
In spite of the small asymmetry between the inner and outer tubes, the disorder-induced inter-tube conductance is essentially determined by the tube distance and remains much smaller than the conductance quantum.
The conductance remains the same even if the relative position of inner and outer lattices is changed, for example, by rotating the inner tube around the tube axis and/or sliding it along the axis, except for a slight change comparable to the differences for the tubes with the same distance shown in Fig.\ \ref{Fig:Various_structures}.
There remain exceptional cases that the inter-tube conductance can become of the order of the conductance quantum, such as in commensurate tubes\cite{Kim_and_Chang_2002a,Tamura_et_al_2005a} and accidental near-commensurate tubes\cite{Uryu_and_Ando_2005c} in the absence of impurities.
Disorder effects in such rare cases are out of the scope of this paper.
\par
%
The role of impurities is to destroy the cancellation of the inter-tube coupling having quasi-periodic position dependence.
This qualitative feature is independent of details of impurities such as their locations, strength, range, etc.
In the case of lattice vacancies (strong, short-range scatterers), the conductance in single-wall tubes is known to depend strongly on their configurations.\cite{Ando_et_al_1999b,Igami_et_al_TB_All}
It is an interesting question whether such strong dependence appears also in the inter-tube conductance of double-wall nanotubes.
Most scatterers in realistic nanotubes are expected to have a potential range larger than the lattice constant.
Such scatterers do not cause backward scattering and cannot contribute to the resistance in metallic single-wall tubes.\cite{Ando_and_Nakanishi_1998a,Ando_et_al_1998b}
The roles of such scatterers in the inter-tube conductance are quite interesting, but are left for a future study.
\par
%
Our results are compared with existing experiments.
From Eq.\ (\ref{Eq:A-linear_normalized_conductance}) with $(t_{\rm max}/\gamma_0)^2\approx0.01$, the inter-tube conductance for short tubes may be given by $G_{41}\sim5\times10^{-3}\times A/l_e$ in units of $e^2/\pi\hbar$.
For the mean-free path $l_e\sim100$ nm,\cite{Schonenberger_et_al_1999a} for example, the inter-tube conductivity is $2G_{41}/A\sim8\ ({\rm m\Omega})^{-1}$ where the factor 2 means the sum of inter-tube transmission and reflection conductances.
This is one-order-of-magnitude smaller than the inter-tube conductivity $\sim100\ ({\rm m\Omega})^{-1}$ estimated in experiments.\cite{Bourlon_et_al_2004a}
\par
%
The calculated conductance of two-terminal tube in Fig.\ \ref{Fig:Two-terminal_conductance} seems to be consistent with the inter-tube conductances raging from $\sim0.03$ to $\sim0.3$ in units of $e^2/\pi\hbar$ obtained for telescoping multi-wall nanotubes.\cite{Cumings_and_Zettle_2004a}
However, details on the disorder remain unknown in the experiments.
Further, much thicker tubes are used in experiments and the actual number of channels may be larger than two,\cite{Bourlon_et_al_2004a} for which inter-tube transfer may be enhanced even in the absence of disorder.\cite{Triozon_et_al_2004a,Roche_et_al_2001c,Ahn_et_al_2003a,Wang_and_Grifoni_2005a}
In this case conduction bands of semiconducting tubes may be populated and the inter-tube conductance of a multi-wall tube including semiconducting tubes is also important, which is closely related to electronic device application.
Clarification of these points is left for future.
\par
%
In inter-tube transport, effects of electron-electron interaction can be important.
The present results of inter-tube coupling are used for the basis for analysis of the interaction effects.
The interaction effects in metallic single- and multi-wall nanotubes were studied theoretically\cite{Wang_and_Grifoni_2005a,Balents_and_Fisher_1997a,Krotov_et_al_1997a,Egger_and_Gogolin_1997a,Kane_et_al_1997a,Yoshioka_and_Odintsov_1999a,Egger_1999a} and experimentally.\cite{Bockrath_et_al_1999a,Yao_et_al_1999a,Ishii_et_al_2003a,Liu_et_al_2001a,Bachtold_et_al_2001a,Graugnard_et_al_2001a,Tarkiainen_et_al_2001a,Kanda_et_al_2004a}
%
If multi-wall tubes are described as weakly coupled Tomonaga-Luttinger liquid\cite{Tomonaga_1950a,Luttinger_1963a} and inter-tube transfer is regarded as tunneling, the inter-tube conductance is likely to exhibit a power-law dependence on the temperature and bias-voltage.
Existing impurities may cause strong backward scattering within each tube,\cite{Voit_1994a} possibly affecting the inter-tube transfer.
Such intriguing problems are left for a future study.
\par
%
\par
%
\section{Summary and conclusion} \label{Sec:Summary_and_conclusion}
%
The inter-tube conductance of double-wall carbon nanotubes with weak short-range impurities is numerically studied.
Length dependence of each of the inter- and intra-tube conductances for various impurities collapses into a single curve when the length is measured in units of the mean-free path.
The inter-tube conductance is substantially enhanced by impurities destroying quasi-periodicity of the system, leading to linear increase with the length.
At length larger than the localization length, the inter-tube transmission  conductance exponentially decays and the inter-tube reflection conductance is saturated.
Their maximum values are much smaller than the conductance quantum.
\par
%
\acknowledgments
%
This work was supported in part by a 21st Century COE Program at Tokyo Tech ``Nanometer-Scale Quantum Physics'' and by Grant-in-Aid for Scientific Research from the Ministry of Education, Culture, Sports, Science and Technology, Japan.
\par
%

%
\end{document}